\documentstyle[epsfig,twoside,fleqn,espcrc2]{article}

\newcommand{\AmS}{{\protect\the\textfont2
  A\kern-.1667em\lower.5ex\hbox{M}\kern-.125emS}}
\title{
\vspace{-8mm}
\rightline{\small HUB--EP--97/58} 
\vspace{-2mm}
\rightline{\small September 9, 1997}
The endpoint of the electroweak phase transition
} 
\author{
  M.~G\"urtler\address{Institut f\"ur Theoretische Physik, Universit\"at
    Leipzig, Augustusplatz 10-11, D-04109 Leipzig, Germany}\thanks{Poster
    presented by M. G\"urtler}, 
  E.-M.~Ilgenfritz\address{Institut f\"ur Physik,
    Humboldt-Universit\"at zu Berlin, Invalidenstr. 110, D-10115 Berlin,
    Germany}, 
  and A.~Schiller$\mathrm{^a}$}
\begin{document}
\begin{abstract}
  The $3d$ $SU(2)$--Higgs model is used to find the critical Higgs mass above
  which the first order phase transition ends.  One method is focused on the
  disappearance of the two--state signal of the scalar condensate (vanishing
  of the latent heat).  Another method is based on the analysis of Lee--Yang
  zeroes of the partition function which allows to characterise the change
  from first order transition into an analytical crossover.
\end{abstract}

\maketitle
\section{Introduction}

Different approaches (see e.g. \cite{buchmueller}) predicted that there is a
critical Higgs mass of $O(100)$ GeV above which there is no first order
thermal phase transition anymore.  The $3d$ $SU(2)$--Higgs model is an
effective model of the electroweak standard model at high temperature and
describes the electroweak phase transition~\cite{generic_rules}.  In this
contribution some results \cite{WIR} obtained within this framework are
summarised to give a more precise characterisation of the end of the phase
transition.

\section{The model}

The lattice action is
\begin{eqnarray}
  S &=& \beta_G \sum_p \left(1 - {1 \over 2}{\mathrm{tr}}
 U_p \right) 
  - \beta_H \sum_{x,\alpha}E_{x,\alpha} \nonumber \\
  && + \sum_x \left( \rho_x^2 + \beta_R \left(\rho_x^2-1\right)^2 \right) \, ,
   \label{eq:latt_action}
\end{eqnarray}
\begin{equation}
  E_{x,\alpha}={1\over 2}{\mathrm {tr}} (\Phi_x^+ U_{x, \alpha} \Phi_{x +
    \alpha}) \, , \  
\rho_x^2=
\frac12{\mathrm {tr}}(\Phi_x^+\Phi_x)    .
  \label{eq:latt_action2}
\end{equation}
An {\it approximate} Higgs mass $M_H^*$ entering
\begin{equation}
  \label{eq:mh*}
  \beta_R=\frac {\lambda_3}{g_3^2}\frac{\beta_H^2}{\beta_G} = \frac
  18{\left(\frac{M_H^*}{80\ {\mathrm {GeV}}}
    \right)}^2\frac{\beta_H^2}{\beta_G}
\end{equation}
is used to label our lattice data.  $\lambda_3$ and $g_3$ are renormalisation
group invariant dimensionful quartic and gauge couplings of the corresponding
$3d$ continuum model.\footnote{The mapping of the lattice couplings to $4d$
  continuum physics is summarised in~\cite{wirNP97}.} The continuum limit
corresponds to $\beta_G \to \infty$.  We use $\rho^2=(1/L^3)\sum_x
\rho_x^2$ as order parameter.

We express lattice sizes and $3d$ results in mass units $g_3^2$ which allows
to combine data obtained at different $\beta_G$. The quality of this scaling
checks to what extent the continuum limit is reached.  In the chosen  units,
the physical lattice size is
\begin{equation}
  \label{eq:phys_units}
  l g_3^2=L a g_3^2= 4L/\beta_G\ .
\end{equation}
The discontinuity of the quadratic scalar condensate is defined by
\begin{equation}
  \label{eq:jump}
  \Delta \left<\phi^+\phi\right>/g_3^2 = 1/8\ \beta_G \beta_H \Delta \left<
    \rho^2 \right>\ 
\end{equation}
through the infinite volume limit of the jump in $\rho^2$ measured at the
pseudo--critical hopping parameter between the broken and symmetric phases.
In our analysis we have widely used the Ferrenberg-Swendsen method where the
reweighting is performed by double--histogramming in two parts of the action.

\section{Latent heat criterion}
\label{sec:latheat}

A first order phase transition is characterised by non--vanishing latent heat
which is proportional to $\Delta \left<\phi^+\phi\right>$. We define the end 
of the  transition line  in the $\beta_H$--$M_H^*$ plane (for
$\beta_G \rightarrow \infty$) where this discontinuity vanishes.  

\begin{figure}[!htb]
  \vspace{-5mm}
  \epsfig{ file=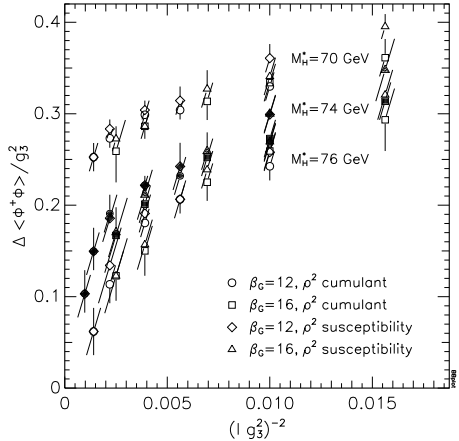, angle=0,width=60mm}
  \vspace{-8mm}
  \caption{Thermodynamical limit for $\Delta\left<\phi^+\phi\right>$ for 
    three values $M_H^*$}
  \label{fig:rho2_extrapol_70+74+76}
  \centering
  \vspace{-5mm}
\end{figure}

The Ferrenberg--Swendsen method allows to  interpolate between different
$\beta_H$ and $M_H^*$.  
In Fig.~\ref{fig:rho2_extrapol_70+74+76} the condensate jump 
for three values of $M_H^*$ and for $\beta_G=12,16$ is plotted as function
of the inverse physical length squared. The respective
pseudo--critical hopping parameters $\beta_{Hc}$ have been defined at the
minimum of the Binder cumulant and maximum of the susceptibility of $\rho^2$.
The plot uses MC data taken in runs at $M_H^*=70, 74, 76$ and $80$~GeV
on cubical lattices.

\begin{figure}[!htb]
  \centering
  \epsfig{file=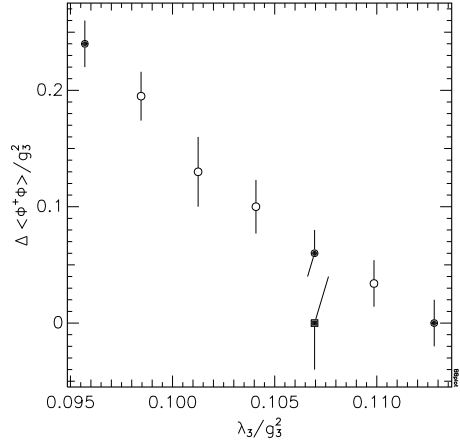,width=60mm}
  \vspace{-1cm}
  \caption{Extrapolation results; circles are extrapolations from at least
    three volumes; the square is an extrapolation from only the two
    largest volumes at that Higgs mass}
  \label{fig:thd_lim}
  \vspace{-5mm}
\end{figure}

Fig.~\ref{fig:thd_lim} represents the infinite volume extrapolations
of $\Delta\left<\phi^+\phi\right>/g_3^2$ as function of $M_H^*$.
The expected quadratic finite size scaling behaviour of
$\Delta\left<\phi^+\phi\right>/g_3^2$ is delayed to lattice sizes not less
than $80^3$ for Higgs masses larger than $70$~GeV. If one includes only the
two largest lattices ($80^3$ and $96^3$ for $M_H^*=74$ GeV) into the
extrapolation the latent heat vanishes for
\begin{equation}
  \label{eq:l3g32_first}
  \lambda_{3\, {\mathrm{crit}}}/g_3^2 < 0.107\  .
\end{equation}

\section{Lee--Yang zero criterion}
A phase transition is characterised by non--analytical behaviour of the
infinite volume free energy density. This is caused by zeroes of the partition
function (extended to complex couplings $\beta_H$) approaching the real axis
in the thermodynamical limit.  At finite volume the zeroes nearest to the real
axis cluster along a line. In the case of a first order phase transition the
first few Lee--Yang zeroes are expected at
\begin{eqnarray}
\label{eq:pattern}
  {\mathrm{Im}} \beta_H^{(n)} &=& 
    \frac{2 \pi \beta_{Hc}}{L^3 (1+2 \beta_{Rc})
    \Delta \left< \rho^2 \right>} \left(n-\frac{1}{2}\right)\\
  {\mathrm{Re}} \beta_H^{(n)}& \approx & \beta_{Hc} \ .
\end{eqnarray}

\begin{figure}[!htb]
  \vspace{-5mm}
  \centering
  \epsfig{file=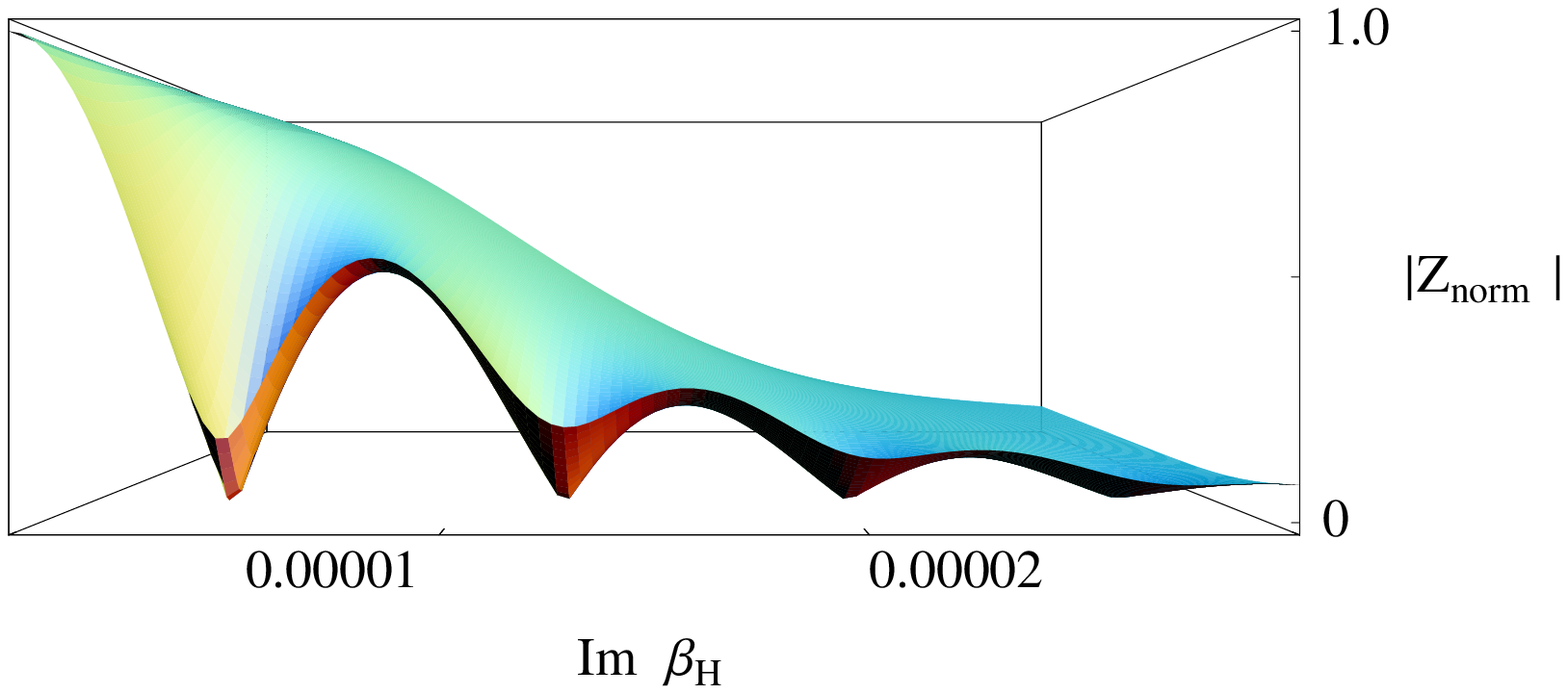,width=60mm,angle=0}
  \vspace{-1cm}
  \caption{$\left|Z_{\mathrm{norm}}\right| =
    \left|Z(\beta_H)/Z({\mathrm{Re}}\, \beta_H)\right|$ near to the first
    four zeroes}
  \label{z2_12_80mod}
  \vspace{-5mm}
\end{figure}

The partition function for complex couplings is obtained by reweighting from
measurements at real couplings. The first zeroes can be well localised
(cf. Fig.~\ref{z2_12_80mod}) 
using the Newton-Raphson algorithm.

\begin{figure}[!htb]
  \centering
  \epsfig{file=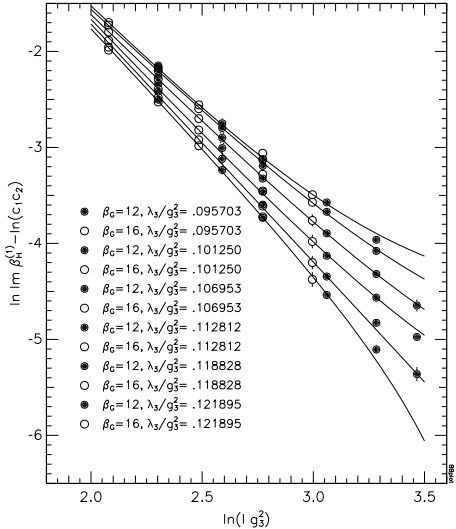, width=60mm,angle=0}
  \vspace{-1cm}
  \caption{
    Logarithm of the imaginary part of first zeroes vs. logarithm of the
    physical length together with the fit (\ref{eq:zero_fit})}
  \label{zero_12_70}
  \vspace{-5mm}
\end{figure}

We fit the imaginary part of the first zero for each available physical length
$lg_3^2$ according to (see Fig.~\ref{zero_12_70})
\begin{equation}
  \label{eq:zero_fit}
  {\mathrm {Im}} \beta_H^{(1)}=C (l\,g_3^2)^{-\nu}+R\ .
\end{equation}
The scenario suggested is the change of the first order transition into an
analytic crossover above ${M_H^*}_c$.  A \emph{positive $R$} signals that the
first zero does not approach anymore the real axis in the thermodynamical
limit.  A similar investigation has been performed recently at smaller gauge
coupling in~\cite{karsch}.

We attempt a global fit according to (\ref{eq:zero_fit}) shifting the zeros as
follows (cf. (\ref{eq:pattern}))
\begin{equation}
  \label{eq:shift_imbetah}
  \ln {\mathrm {Im}}\beta_H^{(1)} \rightarrow \ln {\mathrm {Im}}\beta_H^{(1)}
  \ - \ln (c_1 c_2)
\end{equation}
where $c_1= {\beta_{Hc}^2}/({\beta_G^2 (1+2\beta_{Rc})})$ and $c_2$ is an
extra free constant (numerically between 1.028 and 1.095).
\begin{figure}[!htb]
  \vspace{-5mm}
  \centering
  \epsfig{ file=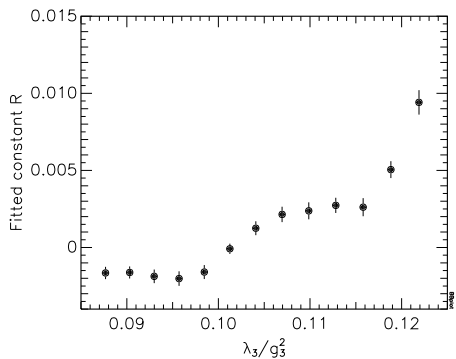, width=60mm,angle=0}
  \vspace{-1cm}
  \caption{ Fitted distance $R$ as function of $\lambda_3/g_3^2$}
  \label{fit_R}
  \vspace{-5mm}
\end{figure}
The fitted $R$  crosses zero (Fig.~\ref{fit_R}) at
\begin{equation}
  \lambda_{3 {\mathrm {crit}}}/g_3^2=0.102(2)\  .
  \label{eq:crit_LY} 
\end{equation}

\section{Conclusions}
\label{sec:conclusions}

We have compared two methods promising to give estimates for the critical
Higgs mass. The criterion based on the correct thermodynamical limit of
Lee--Yang zeroes gives the critical coupling ratio (\ref{eq:crit_LY}). The
vanishing of the discontinuity of the quadratic condensate seems to predict a
somewhat larger critical coupling ratio.  Very near to the critical Higgs mass
one needs data at larger volumes as our analysis has shown.

Using relations between $3d$ and $4d$ quantities we obtain from
(\ref{eq:crit_LY}) the critical (zero temperature) Higgs mass $m_H$ and the
corresponding critical temperature $T_c$ for two versions of the $4d$
$SU(2)$--Higgs theory, without fermions and including the top quark as given
in Table~\ref{tab:latheat} ($g^2(m_W)$ is the renormalised $4d$ gauge
coupling).
\begin{table}[!htb]
  \vspace{-8mm}
  \begin{center}
    \leavevmode
    \begin{tabular}{|c|c|c|}
      \hline
      &&\\[-2ex]
      $m_H/$GeV  & $T_c$/GeV & $g^2(m_W)$ \\[+0.5ex]
      \hline
      67.0(8)  & 154.8(2.6)& 0.423      \\
      72.4(9)  & 110.0(1.5)  & 0.429     \\
      \hline
    \end{tabular}
    \vspace{0.3cm}
    \caption{
      $m_H$ and $T_c$ at $\lambda_{3{\mathrm {crit}}}/g_3^2=0.102$;
      upper row without fermions, lower including top}
    \label{tab:latheat}
  \end{center}
  \vspace{-6mm}
\end{table}

\end{document}